\documentclass[12pt,fleqn]{article}
\begin{document}
\title{\bf Bimaximal mixing from the leptonic new texture for      
triangular mass matrices} 
\author{ H. B. Benaoum\thanks{benaoum@thep.physik.uni-mainz.de} \\
Institut f\"ur Physik, Theoretische Elementarteilchenphysik, \\
Johannes Gutenberg--Universit\"at, \\
55099 Mainz,  Germany
\and 
S. Nasri\thanks{snasri@suhep.phy.syr.edu} \\
Department of Physics, Syracuse University \\
Syracuse, NY 13244--1130, USA }
\date{}
\maketitle
~\\
\abstract{An analysis of the leptonic texture for the new triangular mass 
matrices has been carried out. In particular, it is shown that both 
bimaximal and nearly bimaximal solutions for solar and atmospheric 
neutrino anomalies can be generated within this pattern. We have also derived 
exact and compact parametrization of the leptonic mixing matrix in terms 
of the lepton masses and the parameters $\alpha, \beta'$ and $\delta$. \\
A consistency with the CHOOZ reactor result for $V_m{_{13}}$ and a 
smallness of the Jarlskog's invariant parameter are obtained.
} \\
~\\
~\\
~\\
{ \bf MZ--TH/99-16  \\
SU--4240--699  } 
\newpage
~\\
\section{Introduction :}
One of the most important challenging tasks of present and future experiments 
is to establish whether or not neutrinos have possible rest masses. A clear 
evidence of non--vanishing masses for neutrinos infuences various research 
areas from particle physics, as well as from astrophysics and cosmology. 
Massive neutrinos are regarded as the best candidate for hot dark matter 
and would play a profound role in the formation and stability of our 
universe structure. \\
At present, the recent SuperKamiokande Collaboration~\cite{fuk1}   
observation has provided a strong evidence for neutrino oscillations 
with large mixing as well as non--zero neutrino masses. 
Similar indications in favor of neutrino oscillations comes from the 
results of the atmospheric neutrino experiments ( Kamiokande~\cite{kam}, 
IBM~\cite{ibm}, Soudan 2~\cite{sou} and MACRO~\cite{mac} ). 
Another sign for this proposal comes also from the solar neutrino 
experiments ( Homestake~\cite{hom}, Kamiokande~\cite{kam1}, GALLEX~\cite{gal}, 
SAGE~\cite{sag} and SuperKamiokande~\cite{fuk2} ). \\ 
Let us mention that there is one more possible indication in favor of 
neutrino oscillation, from the Laboratory experiment by the Liquid 
Scintillation Neutrino Detector ( LSND )~\cite{lsnd} at Los Alamos, but 
preliminary data from KARMEN~\cite{karm} failed to reproduce this 
evidence. Here we adopt a conservative approach by not taking into 
account the possibility alluded by the LSND data, waiting for its 
confirmation by other experiments. \\  
A solution to the solar and atmospheric neutrino problem with neutrino 
oscillations requires the existence of two different scales of neutrino 
mass--squared differences, which corresponds to the existence of three 
massive neutrinos. These three massive neutrinos are mixings of three flavor 
neutrinos whose existence is known from the measurements of the invisible 
width of the $Z$ boson done by LEP experiments~\cite{lep}. \\
In the solar neutrino experiments three possible solutions have been proposed 
through the matter enhanced neutrino oscillation (i.e. MSW 
solution~\cite{smi} ) if $\Delta m_{\odot}^2 \simeq 5 \times 10^{-6} eV^2$ 
and $\sin^2 2 \theta_{\odot} \simeq 6 \times 10^{-3}$ ( small angle case ), 
or  $\Delta m_{\odot}^2 \simeq 2 \times 10^{-5} eV^2$ 
and $\sin^2 2 \theta_{\odot} \simeq 0.76$ (large angle case ) and through 
the long--distance vacuum neutrino oscillation called the ''just so'' 
vacuum oscillation $\Delta m_{\odot}^2 \simeq 8 \times 10^{-11} eV^2$ and 
$\sin^2 2 \theta_{\odot} \simeq 0.75$. \\
On the other hand, since the CHOOZ experiment~\cite{cho} excludes 
oscillation of $\nu_{\mu} \rightarrow \nu_e$ with a large mixing angle 
for $\Delta m_{atm}^2 \ge 9 \times 10^{-4} eV^2$, the atmospheric $\nu_{\mu}$ 
deficit is explained by the maximal mixing between $\nu_{\mu}$ and 
$\nu_{\tau}$. \\
Phenomenological analysis favor two solutions for the solar and 
atmospheric neutrino problem, \\ 
~\\
$\ast$~large $\nu_{\mu}--\nu_{\tau}$ mixing for 
the atmospheric anomalies and 
matter enhanced ( MSW ) small mixing angle oscillations for solar neutrinos. \\
~\\
$\ast$~vacuum oscillations and bimaximal or nearly 
bimaximal mixing of three 
light neutrinos. \\
~\\
In the minimal Standard Model based on left--handed two component neutrino 
fields and no right--handed neutrino fields in the Lagrangian, neutrino 
are two component massless particles. A simplest extension of the 
Standard Model with massive neutrinos is obtained by adding right--handed 
neutrino field $\nu_R$ per family with the $SU(2)_L \otimes U(1)_Y$ 
quantum numbers $(0,0)$. Neutrinos acquire then Dirac masses by analogy 
with the quarks and charged leptons. Neutrino mass eigenstates are 
then different from the weak eigenstates, leading to neutrino mixing and 
the violation of family lepton numbers. \\  
~\\
The Yukawa interaction for three generations is
parametrized in terms of $3 \times 3$ matrices which contains a large amount
of free parameters compared to the physical measurables one. \\
To overcome such a freedom, extra symmetries ( ans\"atze ) are introduced
to cast the fermion mass matrices in some particular form~\cite{fri,wil,geo,
ram}. In particular, a general classification and analysis of symmetric
or hermitian mass matrices having textures zeroes consistent with the measured
values of the fermions masses and mixing angles has been carried out
in~\cite{ram}. \\
For non--hermitian mass matrices, Branco et al.~\cite{bra} have shown that
for three generations, fermion masses with textures zeroes can just be obtained
by redefining the fermion fields in a special weak basis transformation which
has no observable consequences. \\
An another class of fermion masses patterns arises in the framework of
Marseille--Mainz noncommutative geometry model~\cite{coq}, namely triangular 
mass matrices~\cite{hau,sch}. They are typical for reducible but indecomposable
representations of graded Lie algebras. \\
It has been shown recently~\cite{hac, ben} that it is possible 
to express these triangular 
mass matrices in a economic and concise way with a minimum set of parameters, 
through a specific weak basis transformation. \\
Indeed, these textures involve 5 complex numbers instead of 6, which means
that one extra parameter is either zero or dependent of the others. A 
connection between these mass matrices and the so--called Nearest--
Neighbor Interactions was established in~\cite{ben}. More details will 
be presented elsewhere~\cite{man}. \\  
~\\
In this article, we would like to examine if such a type of mass matrix 
can be suitable for lepton sector and explore a simple form of triangular 
neutrino mass matrix $T_{\nu}$, which contributes to the $\nu{\mu} 
--\nu_{\tau}$ mixing and a charged triangular lepton mass 
matrix $T_l$,
\begin{eqnarray} 
T_{\nu} & = & \left( \begin{array}{ccc} 
\alpha' & 0 & 0 \\
0 & \beta' & 0 \\
0 & k'_2 & \gamma' \end{array} \right)
\end{eqnarray}
where $\alpha', \beta', \gamma'$ and $k'_2$ are positive parameters and 
\begin{eqnarray}
T_l & = & \left( \begin{array}{ccc}
\alpha & 0 & 0 \\
k_1 e^{i \phi_1} & \beta & 0 \\
0 & k_2 e^{i \phi_r} & \gamma \end{array} \right)~
=~ P^{\dagger} T'_l P 
\end{eqnarray}
with 
\begin{eqnarray} 
T'_l~~=~~ \left( \begin{array}{ccc}
\alpha & 0 & 0 \\
k_1 & \beta & 0 \\
0 & k_2 & \gamma \end{array} \right)~,~~~
P~~=~~ \left( \begin{array}{ccc}
e^{i \phi_1} & 0 & 0 \\
0 & 1 & 0 \\
0 & 0 & e^{- i \phi_r} \end{array} \right) 
\end{eqnarray}
where $\alpha, \beta, \gamma, k_1$ and $k_2$ are all real positive. \\
The characteristic feature of this type of lepton mass matrices is that 
the contribution to $\nu_e--\nu{\mu}$ mixing comes not from the neutrino 
mass matrix $T_{\nu}$ but from the charged lepton $T_l$ and the contribution 
to $\nu_{\mu}--\nu_{\tau}$ mixing angle comes from both $T_{\nu}$ and $T_l$. 
Therefore, we can obtain a large $\nu_{\mu}--\nu_{\tau}$ mixing angle from 
$T_{\nu}$ by taking the small contribution from $T_l$. It will be shown that  
this parametrization is a good candidate pattern to describe the bimaximal and 
nearly bimaximal mixing between $\nu_e--\nu_{\mu}$ and 
$\nu_{\mu}--\nu_{\tau}$. \\  
Our paper is organized as follows. In section 2 we review how to 
transform triangular mass matrices into new triangular forms through 
a weak basis transformation. In particular leptonic mass matrices (1) and 
(2) are obtained when choosing a specific weak basis. In section 3 we generate 
the bimaximal and nearly bimaximal mixing from these patterns and 
reconstruct the mass matrices corresponding to this bimaximal solution. 
In section 4 we end up with some conclusions.
\section{New triangular mass matrix applied to leptons :}
Here we adopt the usual attitude in considering that the nonidentity of the 
mass eigenstates and weak interaction states leads to the weak mixing of 
fermionic states with equal charges but different flavors which means 
that the weak mixing matrix relevant for the charged current interactions 
is obtained from the mass matrices. \\
Moreover, the flavor structure of the Yukawa interactions is not 
constrained by any symmetry but the charged current interactions depend only 
on the left handed fermion fields. Thus, there is much freedom in defining 
a weak basis for the fermions where the full information contained in any 
nonsingular mass matrix ${\cal M} = {\cal T} . {\cal U}_t$, can be recasted by 
means of a triangular mass matrix ${\cal T}$. \\
This type of matrix is obtained from the freedom in choosing the right 
handed basis through ( unobservable ) unitary transformation ${\cal U}_t$ 
and from the fact that the charged current weak interactions involve 
left--handed fields only.  
It has been shown in~\cite{hau} that the triangular mass matrix 
corresponds to the classification of the fermion families in reducible but 
indecomposable representations. \\
~\\
In what follows we consider the mass and the weak charged current Lagrangian 
terms for Dirac neutrinos and charged leptons,
\begin{eqnarray} 
{\cal L} & = & \overline{\nu_L} {\cal T}_{\nu} \nu_R + 
\overline{l_L} {\cal T}_l l_R + 
g~ \overline{\nu_L} \not{W}^+ l_L + h.c. 
\end{eqnarray}
where ${\cal T}_{\nu}$ and ${\cal T}_l$ are lower triangular mass matrices,
\begin{eqnarray}
{\cal T}_{\nu}~=~\left( \begin{array}{ccc}
\alpha' & 0 & 0 \\
k'_1 e^{i \phi'_1} & \beta' & 0 \\
k'_3 e^{i \phi'_3} & k'_2 e^{i \phi'_2} & \gamma' \end{array} \right)~,~~
{\cal T}_l~=~\left( \begin{array}{ccc}
\alpha & 0 & 0 \\
k_1 e^{i \phi_1} & \beta & 0 \\
k_3 e^{i \phi_3} & k_2 e^{i \phi_2} & \gamma \end{array} \right)
\end{eqnarray}
As it is well known, the physics is invariant if the following tranformations 
$\nu_R \rightarrow {\cal V}_{\nu}~\nu_R, l_R \rightarrow {\cal V}_l~l_R, 
\nu_L \rightarrow {\cal U}~\nu_L$ and  $l_L \rightarrow {\cal U}~l_L$ are 
performed on the right and left handed lepton fields where the matrices 
${\cal U}, {\cal V}_{\nu}$ and ${\cal V}_l$ are unitary. This means that 
all sets of mass matrices related to each others through,
\begin{eqnarray}
{\cal T}'_{\nu}~~=~~{\cal U}^{\dagger} {\cal T}_{\nu} {\cal V}_{\nu}~,~~
{\cal T}'_l~~=~~{\cal U}^{\dagger} {\cal T}_l {\cal V}_l
\end{eqnarray}
give rise to the same masses and mixings. \\
In particular, it has been shown recently~\cite{hac, ben} that for any 
arbitrary triangular $3 \times 3$ mass matrices ${\cal T}_{\nu}$ and 
${\cal T}_l$, we can always find weak basis for lepton fields such that the 
hermitian $( {\cal T}'_{\nu} {\cal T}'_{\nu}{^{\dagger}} )_{ij} = 
( {\cal T}'_l {\cal T}'_l{^{\dagger}} )_{ij} = 0$ for fixed $i$ and $j$ such 
that $i \neq j$ where the new triangular mass matrices ${\cal T}'_{\nu}$ 
and ${\cal T}'_l$ are obtained from the above unitary transformations. \\ 
We have also supplied a classification of all possible textures that  
contain the minimal set of parameters and written the original non--hermitian 
mass matrix in terms of the triangular mass matrix elements making 
therefore a bridge and a close connection between the Nearest--
Neighbor Interactions ( NNI ) and the new triangular forms~\cite{ben}. \\
~\\
We will be concerned here, with those textures corresponding to vanishing 
matrix elements $(1,3)$,
\begin{eqnarray}
( {\cal T}'_{\nu} {\cal T}'_{\nu}{^{\dagger}} )_{13} & = & 
( {\cal U}^{\dagger} {\cal T}_{\nu} {\cal T}_{\nu}{^{\dagger}} 
{\cal U} )_{13}~=~ 0 \nonumber \\
( {\cal T}'_l {\cal T}'_l{^{\dagger}} )_{13} & = & 
( {\cal U}^{\dagger} {\cal T}_l {\cal T}_l{^{\dagger}} 
{\cal U} )_{13}~=~ 0 
\end{eqnarray}
where in this basis the new triangular mass matrices ${\cal T}'_{\nu}$ and 
${\cal T}'_l$ are given as follows, see~\cite{hac, ben},
\begin{eqnarray}
{\cal T}'_{\nu}~=~\left( \begin{array}{ccc} 
\alpha' & 0 & 0 \\
k'_1 e^{i \phi'_1} & \beta' & 0 \\
0 & k'_2 e^{i \phi'_2} & \gamma' \end{array} \right)~,~~
{\cal T}'_l~=~\left( \begin{array}{ccc} 
\alpha & 0 & 0 \\
k_1 e^{i \phi_1} & \beta & 0 \\
0 & k_2 e^{i \phi_2} & \gamma \end{array} \right)
\end{eqnarray}
To construct the appropriate unitary matrix ${\cal U}$ that gives this 
requirement, we have first obtained ${\cal U}_{j1}$ which is the 
eigenvectors of the matrix 
$( {\cal T}_{\nu} {\cal T}_{\nu}^{\dagger} + 
k~ {\cal T}_l {\cal T}_l^{\dagger} )$ with $\lambda$ as eigenvalue, 
\begin{eqnarray} 
( {\cal T}_{\nu} {\cal T}_{\nu}^{\dagger} + 
k~ {\cal T}_l {\cal T}_l^{\dagger} )_{ji} {\cal U}_{i1} & = & 
\lambda~ {\cal U}_{j1} 
\end{eqnarray}
where $k$ is a complex parameter expressing the way the neutral and 
charged leptons are correlated. From ${\cal U}_{j1}$, the whole unitary 
matrix ${\cal U}$ can be recovered. \\
We can rewrite this for arbitrary complex $k$ as : 
\begin{eqnarray} 
k~~=~~\frac{\alpha' k'_1}{\alpha k_1} e^{i (\phi_1 - \phi'_1)}~,~~
\lambda~~=~~\alpha'{^2} + k~ \alpha^2
\end{eqnarray} 
In particular the condition,
\begin{eqnarray}
k'_1 & = & 0 
\end{eqnarray} 
accomodates the atmospheric neutrino solution which corresponds to a choice of 
a specific weak basis. \\
The above mass matrices with (11) can be rewritten as :
\begin{eqnarray} 
{\cal T}'_{\nu}~=~\left( \begin{array}{ccc}
\alpha' & 0 & 0 \\
0 & \beta' & 0 \\
0 & k'_2 e^{i \phi'_2} & \gamma' \end{array} \right)~=~
 P_{\nu}^{\dagger} T'_{\nu} P_{\nu},~
{\cal T}'_l~=~\left( \begin{array}{ccc}
\alpha & 0 & 0 \\
k_1 e^{i \phi_1} & \beta & 0 \\
0 & k_2 e^{i \phi_2} & \gamma \end{array} \right)~=~
 P_l^{\dagger} T'_l P_l
\end{eqnarray}
where $T'_{\nu, l}$ are real mass matrices and $P_{\nu, l}$ are diagonal mass 
matrices given by, 
\begin{eqnarray} 
T'_{\nu} & = & \left( \begin{array}{ccc}
\alpha' & 0 & 0 \\
0 & \beta' & 0 \\
0 & k'_2 & \gamma' \end{array} \right)~,~~
P_{\nu}~=~\left( \begin{array}{ccc} 
1 & 0 & 0 \\
0 & 1 & 0 \\
0 & 0 & e^{-i \phi'_2} \end{array} \right)~,  \nonumber \\
T'_l & = & \left( \begin{array}{ccc}
\alpha & 0 & 0 \\
k_1 & \beta & 0 \\
0 & k_2 & \gamma \end{array} \right)~,~~
P_l~=~\left( \begin{array}{ccc} 
e^{i \phi_1} & 0 & 0 \\
0 & 1 & 0 \\
0 & 0 & e^{-i \phi_2} \end{array} \right)
\end{eqnarray}
A particularly remarkable feature of this texture is that the contribution to 
$\nu_e--\nu_{\mu}$ mixing comes from the charged lepton mass matrix $T'_l$ 
which can be taken large in order to describe the "just so " vacuum 
oscillation solution for the solar neutrino problem whereas the neutrino       
$T'_{\nu}$ and the charged lepton $T'_l$ mass matrices contribute to 
$\nu_{\mu}--\nu_{\tau}$ mixing. To explain the atmospheric and solar 
neutrino anomalies,   
it suffices to take a large mixing angle between the second and third 
generations from the neutrino sector and large mixing between the first 
and second generations from the charged lepton sector. \\    
Next, we make the following observation. By using the freedom in choosing 
the right handed fields for both sectors, as well as the left handed 
fields through the neutral diagonal phase matrix $P_{\nu}$,
\begin{eqnarray}
\nu_{L,R} \rightarrow P_{\nu}~\nu_{L,R}~~,~~~~
l_{L,R} \rightarrow P_{\nu}~l_{L,R}
\end{eqnarray}
it leads to a real mass matrix for the neutral sector,
\begin{eqnarray}
T_{\nu} & = & T'_{\nu}~=~\left( \begin{array}{ccc} 
\alpha' & 0 & 0 \\
0 & \beta' & 0 \\
0 & k'_2 & \gamma' \end{array} \right) 
\end{eqnarray}
and a complex mass matrix for the charged sector,
\begin{eqnarray}
T_l & = & ( P_l P_{\nu}^{\dagger} ){^{\dagger}} 
T'_l ( P_l P_{\nu}^{\dagger} )
~=~\left( \begin{array}{ccc}
\alpha & 0 & 0 \\
k_1 e^{i \phi_1} & \beta & 0 \\
0 & k_2 e^{i \phi_r} & \gamma \end{array} \right)
\end{eqnarray}
These lepton mass matrices have 11 parameters, i.e. 4 real moduli 
for neutral sector and 5 real moduli with two phases $\phi_1$ and 
$\phi_r = \phi_2 - \phi'_2$ for the charged sector. \\  
~\\ 
Our purpose is to study the relations between the 
physical quantities and eleven parameters for the neutral $T_{\nu}$ and 
charged $T_l$ lepton mass matrices. \\
Introducing the three neutral lepton mass eigenvalues 
$D_{\nu} = ( m_{{\nu}_e}, - m_{{\nu}_{\mu}}, m_{{\nu}_{\tau}} )$ for 
$T_{\nu}$ as input parameters, $T_{\nu}$ will be parametrized by one 
free parameter. Indeed, the characteristic polynomial for the neutral 
symmetric matrix $T_{\nu} T_{\nu}^{T}$ has the following relations 
between the squared masses and the parameters,
\begin{eqnarray}
m_{{\nu}_e}^2 + m_{{\nu}_{\mu}}^2 + m_{{\nu}_{\tau}}^2 & = & 
\alpha'{^2} + \beta'{^2} + \gamma'{^2} + k'_2{^2} \nonumber \\
m_{{\nu}_e}^2  m_{{\nu}_{\mu}}^2 + m_{{\nu}_e}^2  m_{{\nu}_{\tau}}^2 + 
m_{{\nu}_{\mu}}^2 m_{{\nu}_{\tau}}^2 & = & \alpha'{^2} \beta'{^2} + 
\alpha'{^2} \gamma'{^2} + \beta'{^2} \gamma'{^2} + 
\alpha'{^2} k'_2{^2} \nonumber \\
m_{{\nu}_e}^2 m_{{\nu}_{\mu}}^2 m_{{\nu}_{\tau}}^2 & = & 
\alpha'{^2} \beta'{^2} \gamma'{^2} 
\end{eqnarray}
When solved for $\alpha', \gamma'$ and $k'_2$, these relations give :
\begin{eqnarray}
\alpha'{^2} & = & m_{{\nu}_e}^2 \nonumber \\
\gamma'{^2} & = & \frac{m_{{\nu}_{\mu}}^2 m_{{\nu}_{\tau}}^2}{\beta'{^2}} 
\nonumber \\
k'_2{^2} & = & - \frac{ ( \beta'{^2} - m_{{\nu}_{\mu}}^2 ) 
( \beta'{^2} -  m_{{\nu}_{\tau}}^2 )}{\beta'{^2}} 
\end{eqnarray}
The neutrino mass matrix can be written in terms of the free 
parameter $\beta'$ as : 
\begin{eqnarray}
T_{\nu} & = & \left( \begin{array}{ccc} 
m_{{\nu}_e} & 0 & 0 \\
0 & \beta' & 0 \\
0 & \frac{\sqrt{(\beta'{^2} - m_{{\nu}_{\mu}}^2) 
( m_{{\nu}_{\tau}}^2 - \beta'{^2})}}{\beta'} & 
\frac{m_{{\nu}_{\mu}} m_{{\nu}_{\tau}}}{\beta'} \end{array} \right) 
\end{eqnarray}
$T_{\nu} T_{\nu}^T$ is diagonalized by an orthogonal matrix $O_{\nu}$, 
\begin{eqnarray}
O_{\nu}^T T_{\nu} T_{\nu}^T O_{\nu} & = & D_{\nu}^2 
\end{eqnarray}
where $O_{\nu}$ is given by : 
\begin{eqnarray}
O_{\nu} & = & \left( \begin{array}{ccc} 
1 & 0 & 0 \\
0 & \cos \theta_{\nu} & \sin \theta_{\nu} \\
0 & - \sin \theta_{\nu} & \cos \theta_{\nu} \end{array} \right)
\end{eqnarray}
with
\begin{eqnarray} 
\tan^2 \theta_{\nu} & = & \frac{\beta'{^2} - 
m_{\nu_{\mu}}^2}{m_{\nu_{\tau}}^2 - \beta'{^2}} 
\end{eqnarray}
In the same way, $T_l$ is also parametrized by two real moduli and two 
phases as free parameters by introducing the three charged lepton mass 
eigenvalues $D_l = ( m_e, - m_{\mu}, m_{\tau} )$ as input 
parameters. Indeed, we have :
\begin{eqnarray}
m_e^2 + m_{\mu}^2 + m_{\tau}^2 & = & \alpha^2 + \beta^2 + \gamma^2 + 
k_1^2 + k_2^2   \nonumber \\
m_e^2 m_{\mu}^2 + m_e^2 m_{\tau}^2 + m_{\mu}^2 m_{\tau}^2 & = & 
\alpha^2 \beta^2 + \alpha^2 \gamma^2 + \beta^2 \gamma^2 + 
\alpha^2 k_2^2 + \gamma^2 k_1^2 + k_1^2 k_2^2   \nonumber \\
m_e^2 m_{\mu}^2 m_{\tau}^2 & = & \alpha^2 \beta^2 \gamma^2 
\end{eqnarray}
When solved, we get :
\begin{eqnarray}
k_1^2 & = & \frac{1}{2} ( m_e^2 + m_{\mu}^2 + m_{\tau}^2 - 2 \alpha^2 - 
\beta^2 ) ( 1 - \sqrt{1 - 4 \frac{(\alpha^2 - m_e^2) (\alpha^2 - m_{\mu}^2)   
(\alpha^2 - m_{\tau}^2)}{\alpha^2 (m_e^2 + m_{\mu}^2 + m_{\tau}^2 -
2 \alpha^2 - \beta^2)^2}}~) \nonumber \\ 
k_2^2 & = & \frac{1}{2} ( m_e^2 + m_{\mu}^2 + m_{\tau}^2 - \beta^2 - 2 \gamma^2 )   
( 1 - \sqrt{1 - 4 \frac{(\gamma^2 - m_e^2) (\gamma^2 - m_{\mu}^2)   
(\gamma^2 - m_{\tau}^2)}{\gamma^2 (m_e^2 + m_{\mu}^2 + m_{\tau}^2 - \beta^2 - 
2 \gamma^2)^2}}~
) \nonumber \\
\gamma^2 & = & \frac{m_e^2 m_{\mu}^2 m_{\tau}^2}{\alpha^2 \beta^2}
\end{eqnarray} 
The charged lepton hermitian matrix $T_l T_l^{\dagger}$ is diagonalized by a 
unitary matrix $U^l_L = P . O_l$ and $P = P_l P_{\nu}^{\dagger}$ is a diagonal 
phase matrix, 
\begin{eqnarray}
U^l_L{^{\dagger}} T_l T_l^{\dagger} U^l_L & = & 
O_l^T T'_l T'_l{^T} O_l~~=~~D_l^2 
\end{eqnarray}
Here $O_l$ is an orthogonal matrix which can be written in terms of its 
eigenvectors, 
\begin{eqnarray} 
O_l & = & \left( \begin{array}{ccc}
x_e & x_{\mu} & x_{\tau} \\
y_e & y_{\mu} & y_{\tau} \\
z_e & z_{\mu} & z_{\tau} \end{array} \right) 
\end{eqnarray}
with 
\begin{eqnarray}
\left( \begin{array}{c} 
x_i \\
y_i \\
z_i \end{array} \right) & = & \frac{1}{f_i} \left( \begin{array}{c} 
\alpha \beta k_1 k_2 \\
- \beta k_2 ( \alpha^2 - m_i^2 ) \\
( \alpha^2 - m_i^2 ) ( \beta^2 - m_i^2 ) - m_i^2 k_1^2 \end{array} \right) 
\end{eqnarray}
and the normalization factors $f_i$ are given as follows, 
\begin{eqnarray}
f_i^2 & = & [ ( \alpha^2 - m_i^2) (\beta^2 - m_i^2) - m_i^2 k_1^2 ]~
\prod_{j, j \neq i} ( m_i^2 - m_j^2 ),~~\mbox{for}~ i,j = e, \mu, \tau 
\end{eqnarray}
The lepton flavor mixing matrix $V_m$ connecting the flavor eigenstates to 
the mass eigenstates can be expressed as a product of  
the two matrices $O_{\nu}$ and $U_L^l = P . O_l$ that diagonalize the 
neutrinos and charged lepton mass matrices respectively, 
\begin{eqnarray}
V_m & = & O_{\nu}{^T} . P . O_l 
\end{eqnarray}
\section{Bimaximal mixing from the new triangular lepton mass matrices :}
The interpretation of SuperKamiokande data is compatible with maximal mixing 
between the atmospheric $\nu_{\mu}$ and $\nu_{\tau}$ and the " just so " 
vacuum oscillation solution of the solar neutrino problem. The combination 
of these two possibilities gives the bimaximal mixing matrix,
\begin{eqnarray}
V_m \rfloor_{bi} & = & \left( \begin{array}{ccc} 
\frac{1}{\sqrt{2}} & - \frac{1}{\sqrt{2}} & 0 \\
\frac{1}{2} & \frac{1}{2} &  - \frac{1}{\sqrt{2}} \\
\frac{1}{2} & \frac{1}{2} &  \frac{1}{\sqrt{2}} \end{array} \right) 
\end{eqnarray}
Such a neutrino mixing pattern has received a great deal of 
attention~\cite{gla,tou}. \\ 
In particular, this can be reconstructed from the product 
of two orthogonal matrices,
\begin{eqnarray}
V_m & = & \left( \begin{array}{ccc} 
1 & 0 & 0 \\
0 & \frac{1}{\sqrt{2}} & - \frac{1}{\sqrt{2}} \\
0 & \frac{1}{\sqrt{2}} & \frac{1}{\sqrt{2}} \end{array} \right) 
\left( \begin{array}{ccc} 
\cos \theta_l & \sin \theta_l & 0 \\
- \sin \theta_l & \cos \theta_l & 0 \\
0 & 0 & 1 \end{array} \right)~=~ O_{\nu}{^T} . O_l 
\end{eqnarray} 
where $O_{\nu}$ corresponds here to the maximal mixing between the second and 
the third generations for neutrinos and $O_l$ is parametrized in terms of 
the mixing angle $\theta_l$ expressing the mixing between the first and second 
generations for the charged leptons. 
In this case, the large angle solar neutrino solution is obtained for 
$\cos \theta_l = - \sin \theta_l = \frac{1}{\sqrt{2}}$. \\
~\\
We return now to the analysis of our lepton mass matrices. In accordance 
with the atmospheric neutrino experiments data, the maximal mixing angle 
between the second and the third generations for neutrino is realized in 
our neutrino mass matrix pattern for $\theta_{\nu} = \frac{\pi}{4}$ which 
implies that, 
\begin{eqnarray}
\beta'{^2} & = & \frac{m_{\nu_{\mu}}{^2} + m_{\nu_{\tau}}{^2}}{2}
\end{eqnarray}
The neutrino mass matrix that reproduces this maximal mixing is given by : 
\begin{eqnarray}
T_{\nu} & = & \left( \begin{array}{ccc} 
m_{\nu_e} & 0 & 0 \\
0 & \sqrt{\frac{m^2_{\nu_{\tau}} + m^2_{\nu_{\mu}}}{2}} & 0 \\
0 & \frac{m_{\nu_{\tau}}^2 - m_{\nu_{\mu}}^2}{\sqrt{2~ 
(m^2_{\nu_{\mu}} + m^2_{\nu_{\tau}})}} & \frac{\sqrt{2}~m_{\nu_{\mu}} 
m_{\nu_{\tau}}}{\sqrt{m^2_{\nu_{\mu}} + m^2_{ \nu_{\tau}}}} 
\end{array} \right)
\end{eqnarray}
We want now to determine the charged lepton mass matrix corresponding to the 
"just so" solution for the solar neutrino problem. A typical matrix $O_l$ 
consistent with this vacuum solution is : 
\begin{eqnarray}
O_l & = & \left( \begin{array}{ccc} 
O_{11} \approx \cos \theta_l & O_{12} \approx \sin \theta_l & O_{13} \\
O_{21} \approx - \sin \theta_l & O_{22} \approx \cos \theta_l & O_{23} \\
O_{31} & O_{32} & O_{33} \end{array} \right) 
\end{eqnarray} 
where $|O_{13}|, |O_{23}|, |O_{31}|$ and $|O_{32}|$ are all small 
parameters and $O_{33} \approx 1$. Note that this pattern implies a 
nearly maximal mixing between the first and second generations. \\
This structure is obtained simply by considering $\gamma^2 = m_{\tau}{^2} (1-  
\delta^2)$ where $\delta \ll$ dimensionless parameter   
and expanding (27) and (28) to the first 
order in $\delta$ by using (23) and (24). \\ 
With respect to this analysis, 
the mixing matrix $O_l$ reads in terms of the charged lepton masses and 
the parameters $\alpha$ and $\delta$ as :
\begin{eqnarray}
O_l~=~\left( \begin{array}{ccc} 
\sqrt{\frac{\Delta m^2_{\mu \alpha}}{\Delta m^2_{\mu e}}} (1 + c_{ee} 
\delta ) & \sqrt{\frac{\Delta m^2_{\alpha e}}{\Delta m^2_{\mu e}}} (1 - 
c_{e \mu} \delta ) & c_{e \tau} \delta \\   
- \sqrt{\frac{\Delta m^2_{\alpha e}}{\Delta m^2_{\mu e}}} (1 + c_{\mu e} 
\delta ) & \sqrt{\frac{\Delta m^2_{\mu \alpha}}{\Delta m^2_{\mu e}}} (1 - 
c_{\mu \mu} \delta ) & c_{\mu \tau} \delta \\  
c_{\tau e} \delta & - c_{\tau \mu} \delta & 
\sqrt{\frac{\Delta m^2_{\tau \mu}}{\Delta m^2_{\tau e}}} (1 - c_{\tau \tau} 
\delta )  \end{array} \right) + O\left( \delta^2 \right) 
\end{eqnarray}
The small dimensionless parameters $c_{ij}$ are given as :
\begin{eqnarray}
c_{ee} & = & \frac{m_e^2 m_{\mu}^2~\Delta m^2_{\alpha e}}{2~ 
\Delta m^2_{\tau e}~ ( \alpha^2 \Delta m^2_{\tau \mu} + 
m_e{^2} \Delta m^2_{\mu \alpha} )} \nonumber \\
c_{\mu e} & = & \frac{m_e^2 m_{\mu}^2~\Delta m^2_{\tau \alpha}}{2~ 
\Delta m^2_{\tau e}~ ( \alpha^2 \Delta m^2_{\tau \mu} + 
m_e{^2} \Delta m^2_{\mu \alpha} )} \nonumber \\
c_{\tau e} & = & m_e m_{\mu}~\sqrt{\frac{
\Delta m^2_{\tau \mu} \Delta m^2_{\alpha e}}{\Delta m^2_{\mu e} 
\Delta m^2_{\tau e} ( \alpha^2 \Delta m^2_{\tau \mu} + 
m_e{^2} \Delta m^2_{\mu \alpha} )}} \nonumber \\
c_{e \mu} & = & \frac{m_e^2 m_{\mu}^2~\Delta m^2_{\mu \alpha}}{2~ 
\Delta m^2_{\tau \mu}~( \alpha^2 \Delta m^2_{\tau \mu} +
m_e{^2} \Delta m^2_{\mu \alpha} )} \nonumber \\
c_{\mu \mu} & = &  \frac{m_e^2 m_{\mu}^2~\Delta m^2_{\tau \alpha}}{2~ 
\Delta m^2_{\tau \mu}~ ( \alpha^2 \Delta m^2_{\tau \mu} +
m_e{^2} \Delta m^2_{\mu \alpha} )} \nonumber \\
c_{\tau \mu} & = & m_e m_{\mu}~\sqrt{\frac{
\Delta m^2_{\tau \mu} \Delta m^2_{\mu \alpha}}{\Delta m^2_{\mu e}
\Delta m^2_{\tau e} ( \alpha^2 \Delta m^2_{\tau \mu} +
m_e{^2} \Delta m^2_{\mu \alpha} )}} \nonumber \\
c_{e \tau} & = & \frac{m_e m_{\mu}}{\Delta m^2_{\tau e}}~
\sqrt{\frac{
\Delta m^2_{\alpha e} \Delta m^2_{\mu \alpha}}{
( \alpha^2 \Delta m^2_{\tau \mu} +
m_e{^2} \Delta m^2_{\mu \alpha} )}} \nonumber \\
c_{\mu \tau} & = & \frac{m_e m_{\mu} \Delta m^2_{\tau \alpha}}
{\Delta m^2_{\tau e}}~\frac{1}
{\sqrt{
( \alpha^2 \Delta m^2_{\tau \mu} +
m_e{^2} \Delta m^2_{\mu \alpha} )}} \nonumber \\
c_{\tau \tau} & = & \frac{m_e^2 m_{\mu}^2 ( (m_{\tau}^2 - \alpha^2)^2 + 
(m_{\mu}^2 - \alpha^2) (\alpha^2 - m_e^2) )}{2~(m_{\tau}^2 - m_e^2) 
(m_{\tau}^2 - m_{\tau}^2) ( \alpha^2 \Delta m^2_{\tau \mu} + 
m_e{^2} \Delta m^2_{\mu \alpha} )}
\end{eqnarray}
where we have used the following notation for the mass squared differences 
$\Delta m^2_{ij} = m_i{^2} - m_j{^2}, 
\Delta m^2_{\alpha i} = \alpha^2 - m_i{^2}$ and $\Delta m^2_{i \alpha} = 
m_i{^2} - \alpha^2$ with $i = e, \mu ,\tau$. \\
In the limit $\delta = 0$ with $m_{\tau} \gg m_{\mu}, m_e$ 
(i.e. $k_2 = 0$ ), we obtain the matrix $O_l$,
\begin{eqnarray}
O_l & = & \left( \begin{array}{ccc}
\sqrt{\frac{\Delta m^2_{\mu \alpha}}{\Delta m^2_{\mu e}}} & 
\sqrt{\frac{\Delta m^2_{\alpha e}}{\Delta m^2_{\mu e}}} & 0 \\
- \sqrt{\frac{\Delta m^2_{\alpha e}}{\Delta m^2_{\mu e}}} & 
\sqrt{\frac{\Delta m^2_{\mu \alpha}}{\Delta m^2_{\mu e}}} & 0 \\   
0 & 0 & 1 \end{array} \right)
\end{eqnarray}
which corresponds to just the mixing between the first and second 
generations and a consistency with the large angle solution for the solar 
neutrino anomalie imply :
\begin{eqnarray}
\alpha^2 & = & \frac{m_{\mu}{^2} + m_e{^2}}{2}
\end{eqnarray}
The charged lepton mass matrix for this maximal solution is :
\begin{eqnarray} 
T'_l & = & \left( \begin{array}{ccc} 
\sqrt{\frac{m_e{^2} + m_{\mu}{^2}}{2}} & 0 & 0 \\
\frac{m_{\mu}{^2} - m_e{^2}}{\sqrt{2 (m_e{^2} + m_{\mu}{^2})}} & 
\frac{\sqrt{2} m_e m_{\mu}}{\sqrt{m_e{^2} + m_{\mu}{^2}}} & 0 \\
0 & 0 & m_{\tau} \end{array} \right)
\end{eqnarray}
and in terms of the phase $\phi_1$, this can be written as : 
\begin{eqnarray}
T_l & = & \left( \begin{array}{ccc} 
\sqrt{\frac{m_e{^2} + m_{\mu}{^2}}{2}} & 0 & 0 \\
\frac{m_{\mu}{^2} - m_e{^2}}{\sqrt{2 (m_e{^2} + m_{\mu}{^2})}}~e^{i \phi_1} &
\frac{\sqrt{2} m_e m_{\mu}}{\sqrt{m_e{^2} + m_{\mu}{^2}}} & 0 \\
0 & 0 & m_{\tau} \end{array} \right)
\end{eqnarray}
The above lepton mass matrices (33) and (39) or (40) permit us as desired 
to obtain bimaximal scenario for solar and atmospheric neutrino. They lead 
automatically to a vanishing $V_m{_{13}}$ mixing matrix element which makes 
$\nu_e--\nu_{\mu}$ and $\nu_{\mu}--\nu_{\tau}$ oscillations to be 
effectively a two--channel problem. \\
From our texture, we obtain for the leptonic mixing matrix 
$V_m$ ( see Appendix ), 
\begin{eqnarray}
V_m & = & \left( \begin{array}{ccc}
0.707 + 10^{-10} \delta & 0.707 - 10^{-10} \delta &
7 \times 10^{-7} \delta \\
- 0.5 - 0.0002~ \delta & 0.5 + 0.0002~ \delta &
- 0.706 + 5.7 \times 10^{-8} \delta \\
- 0.5 + 0.0002~ \delta & 0.5 - 0.0002~ \delta &
0.706 - 5.3 \times 10^{-8} \delta \end{array} \right)
\end{eqnarray} 
where we have ignored the phases and used the maximal solutions (32), (38)     
and the following charged lepton masses at the $Z$--boson 
mass scale~\cite{par}, 
\begin{eqnarray} 
m_e (m_Z) = 0.487 MeV~;~~m_{\mu} (m_Z) = 102.7 MeV~;~~
m_{\tau} (m_Z) = 1746.5 \pm 0.3 MeV 
\end{eqnarray} 
This result is in a good agreement with the nonvanishing but small 
$V_m{_{13}}$ coming from the CHOOZ long--baseline neutrino oscillation 
experiment that measures $\overline{\nu_e}$ disapearence. \\
Moreover, the Jarlskog's parameter $J$, which is related to CP violation, is 
given by, 
\begin{eqnarray}
J & = & Im\left( V_m{_{12}} V_m{_{13}}^{\ast}  V_m{_{22}}^{\ast}  
V_m{_{23}} \right) \nonumber \\
& = & \frac{\delta}{4}~\sqrt{\frac{\Delta m^2_{\tau \mu}}
{\Delta m^2_{\tau e}}}~c_{e \tau} \sin 2 \theta_{\nu} \sin 2 \theta_l 
\sin \phi_r + O\left( \delta^2 \right) \nonumber \\
& \approx & 10^{-7}~\delta 
\end{eqnarray} 
showing the smallness of the CP violation phenomena in the leptonic 
sector. 
\section{Conclusion :}
In summary, we have investigated an extension of the Standard Model with 
right handed Dirac neutrino per family by using the new textures for 
fermionic triangular mass matrices. We have proposed new triangular 
mass matrices of the type (8) as the most general description of   
the leptonic sector since the charged current weak interactions involve 
only left--handed fields and the physics is invariant under the weak 
transformations (6). \\
In particular, the patterns (1) and (2) can accomodate the bimaximal and 
nearly bimaximal solutions for atmospheric and solar anomalies. We have 
arrived also at compact formulae for the leptonic mixing matrix in terms 
of the masses and the parameters $\alpha, \beta'$ and $\delta$. \\
Moreover, we have found that they are consistent with the CHOOZ reactor data 
and account for the smallness of the CP violation in the leptonic 
sector. \\
Instead of considering Dirac neutrinos, it is possible to carry out a 
similar analysis for Majorana neutrinos where an extensive use of the 
new triangular mass matrices is taken. Here the Majorana symmetric 
mass matrix is $M_{\nu} = T_{\nu} M_N^{-1} T_{\nu}{^T}$ and the charged lepton 
mass matrix $T_l$ is given by (2). This study is under investigation~
\cite{nas}.
\section*{Acknowledgment} 
We would like to thank Profs. J. Schechter and F. Scheck for reading the 
manuscript and for their valuable comments. \\
One of us (H.B.B.) would like to acknowledge the useful discussions with 
G. Barenboim and the financial support from the DAAD.
\section*{Appendix}
For sake of completeness we give here the expressions for the mixing matrix 
$V_m$ in terms of the lepton masses, $\alpha, \beta'$ and $\delta$, 
\begin{eqnarray*}
V_m{_{ij}} & = & O_{\nu}{_{1i}} . O_l{_{1j}}~e^{i \phi_1} + 
O_{\nu}{_{2i}} . O_l{_{2j}} + O_{\nu}{_{3i}} . O_l{_{3j}}~ 
e^{- i \phi_r}
\end{eqnarray*}
Hence,
\begin{eqnarray*}
V_m{_{11}} & = & \sqrt{\frac{\Delta m^2_{\mu \alpha}}{\Delta m^2_{\mu e}}} 
( 1 + c_{ee}~\delta )~e^{i \phi_1} \\ 
V_m{_{12}} & = & \sqrt{\frac{\Delta m^2_{\alpha e}}{\Delta m^2_{\mu e}}} 
( 1 - c_{e \mu}~\delta )~e^{i \phi_1} \\ 
V_m{_{13}} & = &  c_{e \tau}~\delta~e^{i \phi_1} \\ 
V_m{_{21}} & = & - \sqrt{\frac{\Delta m^2_{\nu_{\tau} \beta'}}
{\Delta m^2_{\nu_{\tau} \nu_{\mu}}}}~
\sqrt{\frac{\Delta m^2_{\alpha e}}{\Delta m^2_{\mu e}}}~(1 + c_{\mu e}~
\delta ) - \sqrt{\frac{\Delta m^2_{\beta' \nu_{\mu}}}
{\Delta m^2_{\nu_{\tau} \nu_{\mu}}}}~c_{\tau e}~\delta~e^{- i \phi_r} \\ 
V_m{_{22}} & = & \sqrt{\frac{\Delta m^2_{\nu_{\tau} \beta'}}
{\Delta m^2_{\nu_{\tau} \nu_{\mu}}}}~
\sqrt{\frac{\Delta m^2_{\mu \alpha}}{\Delta m^2_{\mu e}}}~(1 - c_{\mu \mu}~
\delta ) + \sqrt{\frac{\Delta m^2_{\beta' \nu_{\mu}}}
{\Delta m^2_{\nu_{\tau} \nu_{\mu}}}}~c_{\tau \mu}~\delta~e^{- i \phi_r} \\ 
V_m{_{23}} & = & \sqrt{\frac{\Delta m^2_{\nu_{\tau} \beta'}}
{\Delta m^2_{\nu_{\tau} \nu_{\mu}}}}~ 
c_{\mu \tau}~\delta - \sqrt{\frac{\Delta m^2_{\beta' \nu_{\mu}}}
{\Delta m^2_{\nu_{\tau} \nu_{\mu}}}}~(1 - c_{\tau \tau}~\delta )~
e^{- i \phi_r} \\ 
V_m{_{31}} & = & - \sqrt{\frac{\Delta m^2_{\beta' \nu_{\mu}}}
{\Delta m^2_{\nu_{\tau} \nu_{\mu}}}}~
\sqrt{\frac{\Delta m^2_{\mu \alpha}}{\Delta m^2_{\mu e}}}~(1 + c_{\mu e}~
\delta ) + \sqrt{\frac{\Delta m^2_{\nu_{\tau} \beta'}}
{\Delta m^2_{\nu_{\tau} \nu_{\mu}}}}~c_{\tau e}~\delta~e^{- i \phi_r} \\ 
V_m{_{32}} & = &  \sqrt{\frac{\Delta m^2_{\beta' \nu_{\mu}}}
{\Delta m^2_{\nu_{\tau} \nu_{\mu}}}}~
\sqrt{\frac{\Delta m^2_{\mu \alpha}}
{\Delta m^2_{\mu e}}}~( 1 - c_{\mu \mu}~\delta ) -  
\sqrt{\frac{\Delta m^2_{\nu_{\tau} \beta'}}
{\Delta m^2_{\nu_{\tau} \nu_{\mu}}}}~c_{\tau \mu}~
\delta~e^{- i \phi_r} \\ 
V_m{_{33}} & = &  \sqrt{\frac{\Delta m^2_{\beta' \nu_{\mu}}}
{\Delta m^2_{\nu_{\tau} \nu_{\mu}}}}~c_{\mu \tau}~\delta + 
\sqrt{\frac{\Delta m^2_{\nu_{\tau} \beta'}}
{\Delta m^2_{\nu_{\tau} \nu_{\mu}}}}~
\sqrt{\frac{\Delta m^2_{\tau \mu}}{\Delta m^2_{\tau e}}}~(1 - c_{\tau \tau}~
\delta )~e^{- i \phi_r}  
\end{eqnarray*} 
In addition, the matrix elements of the charged mass matrix 
reconstructed in terms of lepton masses, $\alpha$ and $\delta$ are : 
\begin{eqnarray*}
\beta & = & \frac{m_e m_{\mu}}{\alpha}~\left( 1 + \frac{\delta^2}{2} 
\right) + O\left( \delta^4 \right) \\
\gamma & = & m_{\tau}~\left( 1 + \frac{\delta^2}{2} \right) +  
O\left( \delta^4 \right) \\
k_1 & = & \frac{\sqrt{\Delta m^2_{\mu \alpha} \Delta m^2_{\alpha e}}}{\alpha} 
+ \frac{m_e^2 m_{\mu}^2 \sqrt{\Delta m^2_{\mu \alpha} \Delta m^2_{\alpha e}}}
{2 \alpha (\alpha^2  \Delta m^2_{\tau \mu} + m_e^2 \Delta m^2_{\mu \alpha})}~
\delta^2 + O\left( \delta^3 \right) \\
k_2 & = & \alpha~\sqrt{\frac{\Delta m^2_{\tau e}  \Delta ^2_{\tau \mu}}
{\alpha^2  \Delta m^2_{\tau \mu} + m_e^2 \Delta m^2_{\mu \alpha}}}~\delta 
+ O\left( \delta^3 \right)
\end{eqnarray*} 


\begin{thebibliography}{99}
\bibitem{fuk1}
Y. Fukuda et al., SuperKamiokande Coll., Phys. Rev. Lett. {\bf 81} 
(1998) p.1562; Phys. Lett. {\bf B436} (1998) p.33; Phys. Rev. Lett. 
{\bf 81} (1998) p.1158; Phys. Lett. {\bf B433} (1998) p.9 
\bibitem{kam}
Y. Fukuda et al., SuperKamiokande Coll., Phys. Lett. {\bf B335} 
(1998) p.237
\bibitem{ibm}
R. Becker--Szendy et al., IBM Coll., Nucl. Phys. {\bf B} ( Proc. Suppl.) 
{\bf 38} (1995) p.331
\bibitem{sou}
W.W.M. Allison et al., Soudan Coll., Phys. Lett. {\bf B391} (1997) p.491 
\bibitem{mac}
M. Ambrosio et al., MACRO Coll., Phys. Lett. {\bf B434} (1998) p.451
\bibitem{hom}
B.T. Cleveland et al. Astrophys. J. {\bf 496} (1998) p.505
\bibitem{kam1}
K.S. Hirata et al., Kamiokande Coll., Phys. Rev. Lett. {\bf 77} 
(1996) p.1683
\bibitem{gal}
W. Hampel et al., GALLEX Coll., Phys. Lett. {\bf B388} (1996) 384 
\bibitem{sag}
D.N. Abdurashitov et al., SAGE Coll., Phys. Rev. Lett. {\bf 77} 
(1996) p.4708
\bibitem{fuk2}
Y. Fukuda, SuperKamiokande Coll., Phys. Rev. Lett. {\bf 81} (1998) 
p.1158 
\bibitem{lsnd}
C. Athanassopoulos et al., LSND Coll., Phys. Rev. Lett. {\bf 75} (1995) 
p.2650; Phys. Rev. Lett. {\bf 77} (1996) p.3082; Phys. Rev. {\bf C54} 
(1996) p.2685; nucl--ex/9706006; nucl-ex/9709006; J.E. Hill, 
Phys. Rev. Lett. {\bf 75} (1995) p.2654 
\bibitem{karm}
R. Armbruster et al, KARMEN Coll., Phys. Rev. {\bf C57} (1998) p.3414; 
Phys. Lett. {\bf B423} (1998) p.15; K. Eitel et al., hep--ex/9809007 
\bibitem{lep}
D. Decamp et al., ALEPH Coll., Phys. Lett. {\bf B235} (1990) p.399; 
M.Z. Akrawy et al., OPAL Coll., Phys. Lett. {\bf B240} (1990) p.497; 
P. Abreu et al., DELPHI Coll., Phys. Lett. {\bf B241} (1990) p.435; 
B. Adeva et al., L3 Coll., Phys. Lett. {\bf B249} (1990) p.341 
\bibitem{smi}
L. Wolfenstein, Phys. Rev. {\bf D17} (1978) p. 2369; S.P. Mikheyev and 
A. Yu. Smirnov, Yad. Fiz. {\bf 42} (1985) p.1441; Sov. J. Nucl. Phys. 
{\bf 42} (1985) p.913; Nuovo Cim. {\bf 9C} (1986) p.17
\bibitem{cho}
M. Apollonio et al.; CHOOZ Coll., Phys. Lett. {\bf B420} (1998) p.397
\bibitem{fri}
H. Fritzsch, Phys. Lett. {\bf B73} (1978) p.317;  
Nucl. Phys. {\bf B155} (1979) p.182
\bibitem{wil}
F. Wilczek, A. Zee,  Phys. Rev. Lett. {\bf 42} (1979) p.421 
\bibitem{geo}
H. Georgi, C. Jarlskog, Phys. Lett. {\bf B86} (1979) p.297  
\bibitem{ram}
P. Ramond, R. G. Roberts, G.G. Ross, Nucl. Phys. {\bf B406} (1993) p.19
\bibitem{coq}
R. Coquereaux, G. Esposito--Far\`ese, F. Scheck, Int. J. Mod. Phys. 
{\bf A7} (1992) p.6555 ;   
R. H\"au{\ss}ling, N.A. Papadopoulos, F. Scheck, Phys. Lett. 
{\bf B260} (1991) 
p.125 ;   
R. Coquereaux, R. H\"au{\ss}ling, N.A. Papadopoulos, F. Scheck, 
Int. J. Mod. Phys. {\bf A7} (1992) p.2809 
\bibitem{hau}
R. H\"au{\ss}ling, F. Scheck, Phys. Lett. {\bf B336} (1994) p.477; \\
See also, R. H\"au{\ss}ling, M. Paschke and  F. Scheck, Phys. Lett. 
{\bf B417} (1998) p.312 
\bibitem{bra}
Branco et al., Phys. Rev. {\bf D39} (1989) p.3443; \\ 
see also, K. Harayama and N. Okamura, Phys. Lett. {\bf B387} (1996) p.614; 
T. Ito, Prog. Theor. Phys. {\bf 96} (1996) p.1055; E. Takasugi, 
Theor. Phys. {\bf 98} (1997) p.177; Y Koide, Mod. Phys. Lett. {\bf A12} 
(1997) p.2655
\bibitem{hac}
H.B. Benaoum, " More on triangular mass matrices for fermions " \\
submitted to Phys. Lett. B  
\bibitem{ben}
H.B. Benaoum, F. Scheck, " New textures for triangular mass matrices " \\
,{\bf MZ--TH/98--36}; Mainz--preprint 5 September 1998.
\bibitem{sch}
R. H\"au{\ss}ling,  F. Scheck,     Phys. Rev. {\bf D57 } (1998) p.6656 
\bibitem{man} 
H.B. Benaoum, R. H\"au{\ss}ling and F. Scheck, work in preparation.
\bibitem{gla}
H. Georgi and S.L. Glashow, hep--ph/9808293 
\bibitem{tou}
V. Barger, S. Pasvasa, T.J. Weiler and K. Whisnant, Phys. Lett. {\bf B437} 
(1998) p.107; A.J. Baltz, A.S. Goldhaber and M. Goldhaber, Phys. Rev. 
Lett. {\bf 81} (1998) p.5730; Y. Nomura and T. Yanagida, Phys. Rev. 
{\bf D59} (1998) p.017303; G. Altarelli and F. Feruglio, Phys. Lett. 
{\bf B439} (1998) p.112; S.M. Bilenky and G. Giunti, hep--ph/9802201;   
E. Ma, hep--ph/9807386; N. Haba, Phys. Rev. 
{\bf D59} (1999) p.035011; H. Fritzsch and Z.Z. Xing, Phys. Lett. 
{\bf B440} (1998) p.313; S. Davidson and S.F. King, hep--ph/9808296; 
R.N. Mohapatra and N. Nussinov, hep--ph/9808301; hep--ph/9809415; 
G. Altarelli and F. Feruglio, hep--ph/9809596; C. Jarlskog, M. Matsuda, 
S. Skadhauge and M. Tanimoto, hep--ph/9812282; A. Ghosal, hep-ph/9903497; 
M. Jezabek and Y. Sumino, hep-ph/9904382 
\bibitem{par} 
Particle Data Group, Phys. Rev. {\bf D54} (1996) p.1; The European 
Physical Journal {\bf C3} (1998) p.1; http://pdg.lbl.gov; B. Stech, 
Phys. Lett. {\bf B403} (1997) p.114 
\bibitem{nas} 
H.B. Benaoum and S. Nasri, work in progress 
\end{thebibliography}
\end{document}